\documentclass[twocolumn,showpacs,preprintnumbers,amsmath,amssymb]{revtex4}

\usepackage{graphicx}
\usepackage{dcolumn}
\usepackage{bm}

\begin{document}

\title{The Evolution of Interpersonal Relationships in a Social Group}
\author{Bo Hu}
\author{Xin-Yu Jiang}
\author{Bing-Hong Wang}
\author{Jun-Feng Ding}
\author{Tao Zhou}
\author{Yan-Bo Xie}
\email{bhwang@ustc.edu.cn, Fax:+86-551-3603574.}
\affiliation{%
Nonlinear Science Center and Department of Modern Physics,
University of Science and Technology of China, Hefei, 230026, PR
China \\
}%

\date{\today}

\begin{abstract}
A model has been proposed to simulate the evolution of
interpersonal relationships in a social group. The small social community
is simply assumed as an undirected and weighted graph, where
individuals are denoted by vertices, and the extent of favor or
disfavor between them are represented by the corresponding edge
weight. One could further define the strength of vertices to
describe the individual popularity. The strength evolution exhibits
a nonlinear behavior. Meanwhile, various acute perturbations to
the system have been investigated. In the framework of
our model, it is also interesting to study the adaptive
process of a new student joining a class midway.

\end{abstract}


\maketitle

\section{Introduction}
A social network is a set of people with some pattern of contacts
or interactions between them\cite{ref1,ref2}. Cases that have been
studied include the patterns of friendships between
individuals\cite{ref3,ref4}, business relationships between
corporations\cite{ref5,ref6}, and intermarriages between
families\cite{ref7}. In a social group, all individuals will have
their friends and enemies. Rapoport and others have once studied
friendship networks of school children\cite{ref4,ref14}, due to
its relatively simple pattern and small sample size. Recently,
Alain Barrat, et al. have proposed a general model for the growth
of weighted networks\cite{ref26}, considering the effect of the
coupling between topology and weights' dynamics. It appears that
there is a need for a modelling approach to complex networks that
goes beyond the purely topological point. We have proposed a simple
model to simulate the interpersonal relationships among pupils of a
class\cite{ref8}. In accordance with empirical observations,
similar human relations between people will promote friendship,
while opposite interpersonal relations may lead to hostility. The
first impression effect, as shown in \cite{ref8}, has played a
significant role in the evolution of the interpersonal relations.
As a further research and development of our original work, this
paper, with both analytical and experimental methods, gives a
detailed analysis to the evolution behaviors. The reactions of the
system to various perturbations are also investigated.

\section{Review The Model}
The model system comprises $N$ students and we introduce a
$N\times N$ weighted adjacency matrix to describe its
interpersonal relationships. The symmetric matrix elements $\omega_{ij}$
represent the weight of edge $e_{ij}$, where $i$, $j$=1, 2,
\ldots, $N$. The value of $\omega_{ij}$ is discrete and can be
negative for the cases of disfavor relationships; each contact
between individuals are supposed to alter weight by $\pm1$ at
most. This assumption makes the pairwise interaction moderate;
thus, love or hatred in this model is not formed in seldom
contacts. In the original model, we assign
value 1 with probability $p$, and -1 with probability 1-$p$ to the
elements of matrix for initialization, where $p$ is called
\emph{the initial amity possibility}. The definition of the model
is based on the weights' dynamics:

(i)First, suppose student $i$ has been randomly selected from the
class. Then, he takes the initiative in contacting student $j$
with possibility:

\begin{equation}
W_{i\rightarrow j}=\frac{|\omega_{ij}|+1}{\sum_{j=1,j\neq
i}^N(|\omega_{ij}|+1)}.
\end{equation}

Obviously,
\begin{equation}
W_{i\rightarrow j}=W_{j\rightarrow i}
\end{equation}

The reason to choose this possibility is to avoid the scenario
that student $j$ with $\omega_{ij}$=0 cannot be selected by
student $i$, (for more information, please see Ref\cite{ref8}).

(ii)Now, $i$ and $j$ have been chosen for interaction. Since
``birds of a feather flock together" and similar human relations
and social environments are likely to promote friendship, we could
define $\gamma_{ij}$ as below to describe the interpersonal
relation similarity:

\begin{equation}
\gamma_{ij}=C^{-1}\sum_{\alpha}\omega_{i\alpha}\cdot\omega_{\alpha
j}
\end{equation}

where
\begin{equation}
C=\sqrt{\sum_{\alpha}\omega_{i\alpha}^{2}}\cdot\sqrt{\sum_{\beta}\omega_{j\beta}^{2}}
\end{equation}

Apparently, $\gamma_{ij}$=$\gamma_{ji}$ and
$-1\leq\gamma_{ij}\leq$1.

Then, the rules of revolution are (for more details, please see
our original work\cite{ref8}):

when $\gamma_{ij}\geq$0
\begin{eqnarray}
\omega_{ij}\rightarrow\omega_{ij}+1,
\omega_{ji}\rightarrow\omega_{ji}+1
\end{eqnarray}
with possibility $\gamma_{ij}$, and nothing is altered with
possibility $1-\gamma_{ij}$;

when $\gamma_{ij}<$0
\begin{eqnarray}
\omega_{ij}\rightarrow\omega_{ij}-1,
\omega_{ji}\rightarrow\omega_{ji}-1
\end{eqnarray}
with possibility $|\gamma_{ij}|$, and nothing is altered with
possibility 1$-|\gamma_{ij}|$.

After the weights have been updated (*the symmetry requirement of
the matrix must be satisfied in every update), the process is
iterated by randomly selecting a new individual for the next
contact until the class disbands.

The most commonly used topological information about vertices is
their degree and is defined as the number of neighbors. A natural
generalization in the case of weighted networks is the strength.
Define strength $\ s_i$ of node $i$ as below to describe individual
popularity:

\begin{equation}
s_i=\sum_{j=1}^N\omega_{ij}.
\end{equation}

We could write the evolution equation of strength $\ s_i$ as
bellow:

\begin{eqnarray}
\frac{\partial s_i}{\partial t} & = & \sum_{j=1,j\neq
i}^{N}\frac{1}{N}W_{i\rightarrow j}\gamma_{ij}+\sum_{j=1,j\neq
i}^{N}\frac{1}{N}W_{j\rightarrow i}\gamma_{ji}
\end{eqnarray}

The first term on the right hand side of Eq. (8) describes all the
possible alterations to $s_i$ at time step $t$, contributed by the
cases that $i$ is first randomly chosen and then takes
initiative in contacting $j$; likewise, the second term describes
all contributions from $j$ choosing $i$. Given that
$W_{i\rightarrow j}$=$W_{j\rightarrow i}$ and
$\gamma_{ij}$=$\gamma_{ji}$, we could simplify master Eq. (8):

\begin{eqnarray}
\frac{\partial s_i}{\partial t} & = & \frac{2}{N}\sum_{j=1,j\neq
i}^{N}W_{i\rightarrow j}\gamma_{ij}\\
              & = & \frac{2}{N}\frac{\sum_{j=1,j\neq i}^{N}(|\omega_{ij}|+1)\gamma_{ij}}{\sum_{j=1,j\neq
              i}^N(|\omega_{ij}|+1)}.
\end{eqnarray}

\begin{figure}
\scalebox{0.8}[0.8]{\includegraphics{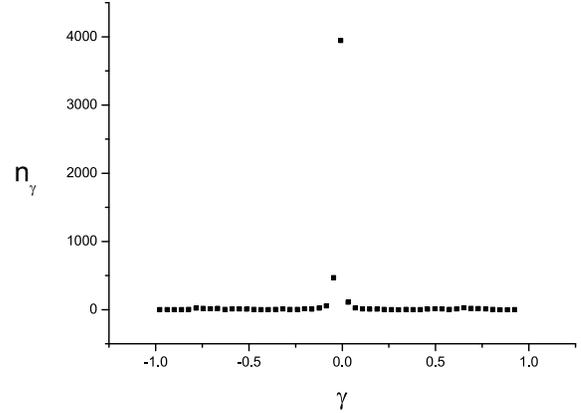}}
\caption{\label{s:p}$\gamma$ distribution for p=0.00, N=100 and
M=50 after 1.0$\times10^{6}$ time steps.}
\end{figure}

\begin{figure}
\scalebox{0.8}[0.8]{\includegraphics{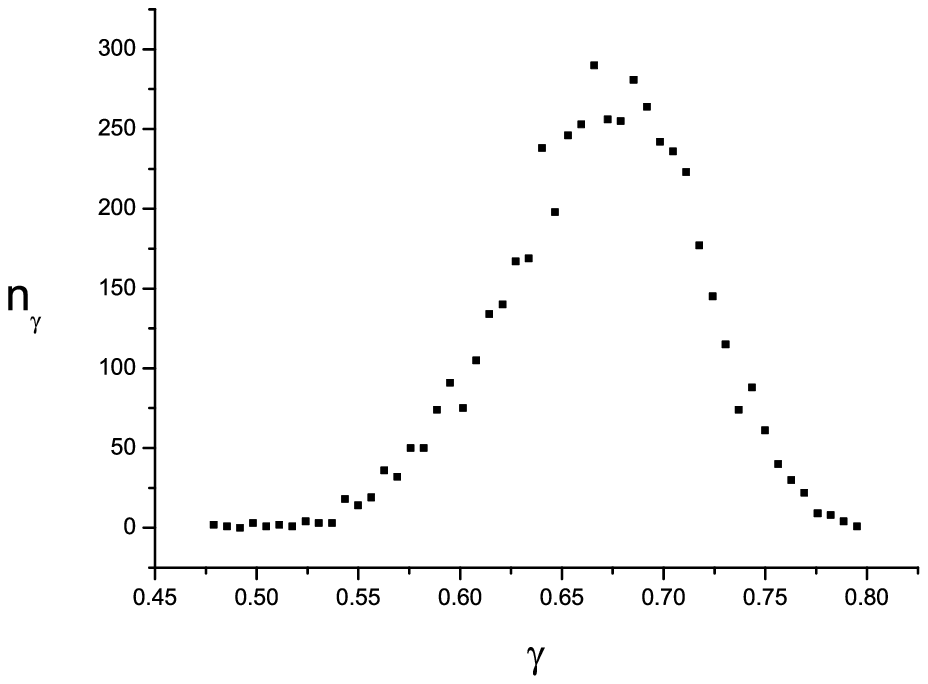}}
\caption{\label{fig:epsart} $\gamma$ distribution for p=1.00, N=100 and M=50 after 1.0$\times10^{6}$ time steps.}
\end{figure}

Weight $\omega_{ij}$ has a certain distribution in the final state
(as shown in Ref.\cite{ref8}), so do $W_{i\rightarrow j}$ and $\gamma_{ij}$.
For convenience, we define:

\begin{equation}
<\gamma>_i=\frac{\sum_{j=1,j\neq i}^{N}(|\omega_{ij}|+1)\gamma_{ij}}{\sum_{j=1,j\neq i}^N(|\omega_{ij}|+1)},
\end{equation}
obviously, $|<\gamma>_i|<$1.

To obtain the distribution of $\gamma$, the range
of $\gamma$ is equally divided by $M$. Then, the range
[$\gamma_{min}$, $\gamma_{max}$] becomes [$\gamma_{1}$,
$\gamma_{2}$), [$\gamma_{2}$, $\gamma_{3}$), \ldots,
[$\gamma_{M}$, $\gamma_{M+1}$], where $\gamma_{1}$=$\gamma_{min}$,
$\gamma_{M+1}$=$\gamma_{max}$. Define $n_{\gamma_{l}}$ as the
number of $\gamma$ between [$\gamma_{l}$, $\gamma_{l+1})$, $l$=1,
2, $\ldots$, $M$; when $l$=$M$ the interval is $[\gamma_{M},
\gamma_{M+1}]$. As shown in Fig. 1, the distribution of $\gamma$
has a pinnacle at $\gamma$=0. The average value of
$\gamma$ is $-$0.00118 as calculated. Since now the weight distribution
also behaves a pinnacle near $\omega$=0\cite{ref8}, it indicates that when the initial
non-diagonal elements(weights) are all $-$1, the system behaves a tardy evolution.
Many contacts will give no alteration to the weights, or equally, the effective
contacts between people that change their relations occur infrequently.
From Fig. 2, one can see that the $\gamma$ distribution exhibits a convexity and reaches the maximum
near $\gamma$=0.67. $\gamma>$0.47 and the average of $\gamma$: $<\gamma>$=0.66817.
Since the weight distribution $n_{\omega}\sim\omega$ also displays
a convexity at $\omega>$0, the effective contacts between individuals are
quite frequent. Above results would be reviewed in the next section.

\section{Perturbations and Evolutions}
In the original model, we suppose that individuals
at first are not familiar with each other.
The initial configuration of the system
is determined by the initial amity possibility
$p$. It can be interpreted as the first impression among them.
However, it is completely possible that two individuals have
good fellowship or a deep grievance previously.
Plotted in Fig. 3 is $\omega_{0}\sim t$ for the process of an old grievance
thawing in an initial harmonious group ($p\approx$1.00). $\omega_{0}$, which
denotes the old grievance between two from the very beginning,
rises sharply from -100 to 0, and increases gently afterwards.
This phenomenon can be interpreted from two aspects. On the one hand,
the new environment provides them sufficient opportunities to contact other people,
so they would not encounter frequently and aggravate their greivance.
On the other hand, they both will get on well quickly with new individuals,
due to $p\approx$1.00 (see Ref.\cite{ref8}). Thus, after sufficient contacts
they will have common friends who will play a conciliatory
role in mitigating their old grievance.

\begin{figure}
\scalebox{0.8}[0.8]{\includegraphics{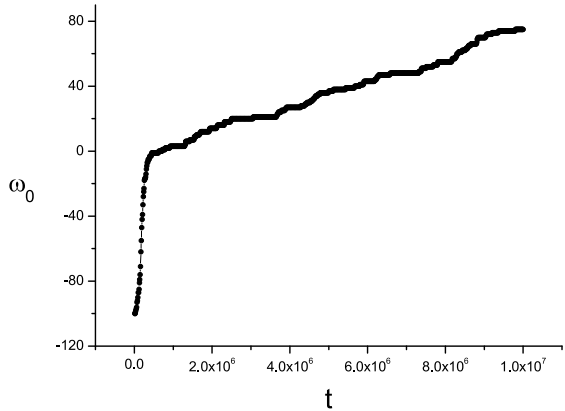}}
\caption{\label{fig:epsart} weight evolution for N=100 after 1.0$\times10^{7}$ time steps.
Initially, all the weights are set +1, except $\omega_{0}$=-100, where $\omega_{0}$
denotes the designated negative weight.}
\end{figure}

All too often, a group has its leader who has great
reverence among his people. He may be an elder chief, a captain
of a football team or a chairman of an association. The other members of
the group may have some conflicts and grievances. We are interested in the
role of the leader playing in the development of the collective.
For simplicity, we could assign a large integer to the matrix elements
at row $i$ ($\omega_{ij}$=$\omega_{ji}$ and $\omega_{ii}$=0) to
denote the leader's relations, and set all the other non-diagonal
elements as -1. In Fig. 4, we have taken this integer as 100. Apparently, the leader's strength
grows linearly with the passage of time, while the strength of his members increases with
a relatively slow velocity. By comparison with the evolution containing no leader, one
could conclude that the strengths of the system have been boosted up by the leader. The slope of the
leader's popularity evolution is 0.03167. In section 2, we have given that:
\begin{equation}
\frac{\partial s_i}{\partial t}=2<\gamma>_i/N.
\end{equation}
For the case containing the leader: if we assume that $\omega_{kj}\ll\omega_{ij}$ for all $j$ and $i\neq k$,
then $\gamma_{ij}\approx$1, and further $\partial s_i/\partial t$=2/$N$=0.04, in approximate agreement
with the experimental results. For the case absent the leader:
the average speed of their popularity growth can be also estimated.
If we suppose $<\gamma>_i\approx<\gamma>$=$-$0.0118 (as mentioned in section 2),
then Eq. (12) gives the average velocity of popularity growth $-$7.2$\times10^{4}$,
which is quite low. But as one can see in Fig. 4, their strengths are dispersed
at a certain time (nonlinear property). Note that the master
equation (8) can only give an average rate of popularity growth.
The leader has changed the possible state of disunity and make his men draw together.

\begin{figure}
\scalebox{0.8}[0.8]{\includegraphics{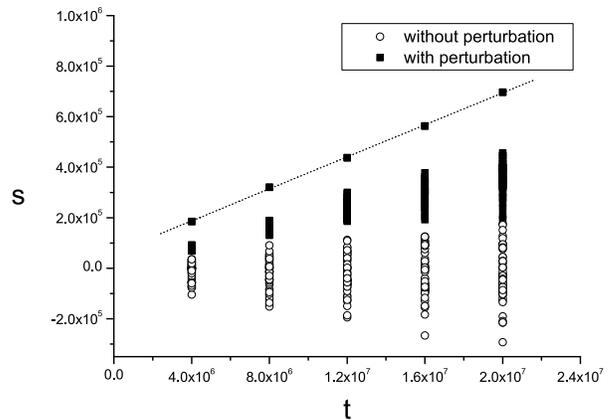}}
\caption{\label{fig:epsart} strength evolution for N=50, M=50 and p=0.00 after 2.0$\times10^{7}$ time steps.
The dot line (with slop 0.03167) fits the strength evolution of the leader, and
blacksquare($\blacksquare$) represents the strength of
others, circle($\circ$) denotes the strength evolution without the leader, under the same conditions. The data
are drawn out at intervals of 4.0$\times10^{6}$ time steps.}
\end{figure}
It is also interesting to study the opposite case. We could put a public
anti into a united group to see the strength evolution of the system, see Fig. 5.
For initialization, we assign $-$100 to the matrix elements at
row $i$ ($\omega_{ij}$=$\omega_{ji}$ and $\omega_{ii}$=0) to represent the anti's relations,
and set all the other non-diagonal elements as 1 to describe the initial public.
Obviously, the anti's popularity decreases linearly,
while the strength of the public rises at a relatively slow speed
(approximate linear). By comparing with the case with no anti,
one could see that the strength evolution
of the public has not been changed significantly by the anti.
The anti's incompatibility to the public is aggravated.
Recall that $<\gamma>$=0.67 in section 2, and suppose $<\gamma>_j\approx<\gamma>$ (for $j\neq i$).
We could estimate the growth rate of the public strength
as $2<\gamma>_j/N\approx2<\gamma>/N$=$2\times0.76/50$=0.0304.
The slope of the anti's popularity evolution is $-$0.0425.
It is reasonable to assume that $|\omega_{kj}|\ll|\omega_{ij}|$ for all $j$ and $i\neq k$,
then $\gamma_{ij}\approx-$1, and further $\partial s_i/\partial t$=2/$N$=$-$0.04,
in fine agreement with the experimental results.

\begin{figure}
\scalebox{0.8}[0.8]{\includegraphics{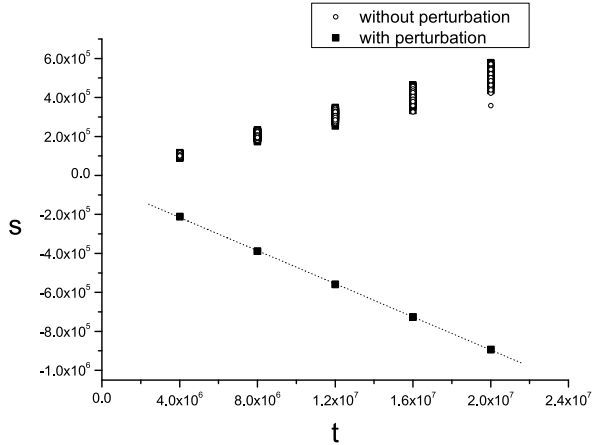}}
\caption{\label{fig:epsart} strength evolution for N=50, M=50 and p=1.00 after 2.0$\times10^{7}$ time steps.
The dot line (with slop -0.0425) fits the strength evolution of the public anti, and
blacksquare($\blacksquare$) represents
the strength of the public, circle($\circ$) denotes the strength evolution without the anti, under the same
conditions. The data are drawn out at intervals of 4.0$\times10^{6}$ time steps.}
\end{figure}

It must be stressed that the strength growth is actually
nonlinear. As Fig. 6 indicates, the trace of the strength
evolution behaves linear initially, but nonlinear afterwards.
Experiments demonstrate that the strength evolution exhibits
obvious nonlinear behaviors after sufficient time, whatever the
initial state. When $p$=1.00, it has a widest linear zone, but
finally, it splits. We would like to discuss the adaptive process
of a student joining a class midway. Plotted in Fig. 8 (log-log
scale) is the strength evolution of the system on the background
of $p$=1.00. The old students' popularity grows linearly, even
after inserting the new one. One could see the nonlinear growth
for the new-coming pupil, whose popularity grows approaching the
asymptote of the old students' strength evolution. The adaptive
process for the new is quite short, thanks to the harmonious
group.
\begin{figure}
\resizebox{0.95\columnwidth}{!}{\includegraphics{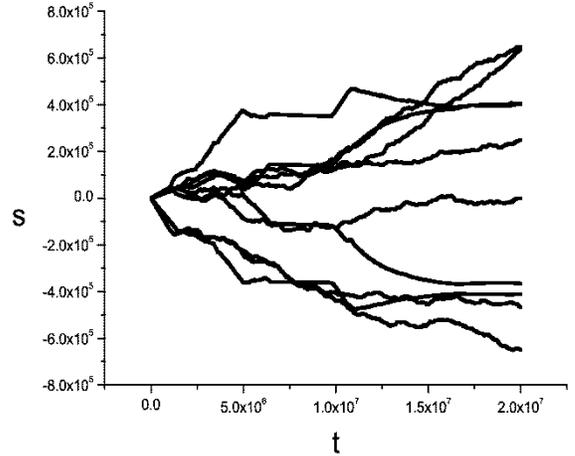}}
\caption{\label{fig:epsart} strength evolution for N=10, p=0.00 after 2.0$\times10^{7}$ time steps.}
\end{figure}

\begin{figure}
\scalebox{0.7}[0.7]{\includegraphics{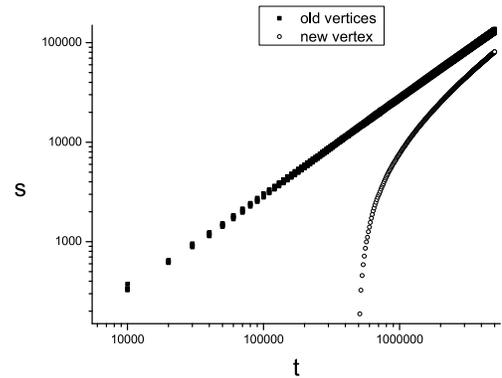}}
\caption{\label{fig:epsart} strength evolution of a student joining midway: N=49 and p=1.00. The new student
joins the class after 4.0$\times10^{5}$ time steps, and his evolution is denoted by $\diamondsuit$.
the strength evolution of old students are denoted by $\blacksquare$. We have extracted 5 from them as
representatives.}
\end{figure}

\section{Outlook and Applications}
This paper is focused on discussing various perturbations and corresponding evolutions
of our original model, which is proposed to study the student relationships
in a class. Both experimental and analytical methods have been used to learn human relations in a
local group. The generalization of the model to describe the real social networks
is still a challenging problem.
(to be continued.)

\begin{acknowledgements}
This work was supported by the State Key Development Programme of
Basic Research of China (973 Project), the National Natural
Science Foundation of China under Grant No.70271070, the
China-Canada University Industry Partnership Program (CCUIPP-NSFC
No.70142005), and the Doctoral Fund from the Ministry of Education
of China.
\end{acknowledgements}


\begin{thebibliography}{ref1}
\bibitem{ref1} Scott, J., {\it Social Network Analysis: A Handbook}, Sage Publications, London,
2nd ed. (2000).

\bibitem{ref2} Wasserman, S. and Faust, K., {\it Social Network Analysis}, Cambrige University
 Press, Cambridge (1994).

\bibitem{ref3} Moreno, J. L., {\it Who Shall Survive?}, Beacon House, Beacon, NY (1934).

\bibitem{ref4} Rapoport, A., and Horvath, W. J., A study of a large
socialgram, {\it Behavior Science 6}, 279-291 (1961).

\bibitem{ref5} Mariolis, P., Interlocking directorates and
control of corporations: The theory of bank control, {\it Social
Science Quarterly 56}, 425-439 (1975).

\bibitem{ref6} Mizruchi, M. S., {\it The American Corporate Network,
1904-1974}, Sage, Beverly Hills (1982).

\bibitem{ref7} Padgett, J. F. and Ansell, C. K., Robust action and
the rise of the Medici, 1400-1434, {\it Am. J. Sociol. 98},
1259-1319 (1993).

\bibitem{ref8} B. Hu, X.-Y. Jiang, J.-F. Ding, Y.-B. Xie. and
B.-H. Wang, Preprint, cond-mat/0408125.


\end{thebibliography}
\end{document}